\documentclass[conference]{IEEEtran}
  \IEEEoverridecommandlockouts%
  \usepackage{cite}
  \usepackage{amsmath,amssymb,amsfonts}
  \usepackage{algorithmic}
  \usepackage{graphicx}
  \usepackage{xcolor}
  \usepackage{textcomp}
  \usepackage{booktabs}
  \usepackage{multirow}
  \usepackage{tabularx}
  \usepackage{siunitx}
  \usepackage[outdir=./Figures/]{epstopdf}
  \DeclareSIUnit{\belmilliwatt}{Bm}
\DeclareSIUnit{\dBm}{\deci\belmilliwatt}
  \usepackage{url}

\usepackage{import}
  \usepackage{caption}
  \usepackage{subcaption}
  \usepackage{comment}
  \usepackage{nccmath} % reduce size of eq.'s with \medmath

  \usepackage{placeins} % for floatbarrier
  \PassOptionsToPackage{hyphens}{url}\usepackage{hyperref}
  
  \usepackage[utf8]{inputenc}
  \usepackage{pgfplots}
  \pgfplotsset{compat=newest}
\pgfplotsset{every axis/.append style={
      line width=0.75pt,
      mark size=1pt,
      axis line style=thin,
      label style={font=\footnotesize},
      legend style={font=\scriptsize, draw=none},
      tick label style={font=\footnotesize},
      axis x line*={bottom},
      axis y line*={left}
    }
}
  \usepgfplotslibrary{groupplots}
  
  \usepackage{mathtools}
  \DeclarePairedDelimiter{\ceil}{\lceil}{\rceil}
  
  \makeatletter
  \providecommand\add@text{}
  \newcommand\tagaddtext[1]{%
  \gdef\add@text{#1\gdef\add@text{}}}%
  \renewcommand\tagform@[1]{%
  \maketag@@@{\llap{\add@text\quad}(\ignorespaces#1\unskip\@@italiccorr)}%
  }
  \makeatother
  
  \def\BibTeX{{\rm B\kern-.05em{\sc i\kern-.025em b}\kern-.08em
  T\kern-.1667em\lower.7ex\hbox{E}\kern-.125emX}}
\begin{document}\sloppy

\title{Cross-Layer Framework and Optimization for Efficient Use of the Energy Budget of IoT Nodes}

\author{\IEEEauthorblockN{Gilles Callebaut, Geoffrey Ottoy, Liesbet Van der Perre}
	\IEEEauthorblockA{
		\textit{KU Leuven, ESAT-DRAMCO, Ghent Technology Campus}\\
		Ghent, Belgium\\
		gilles.callebaut@kuleuven.be, geoffrey.ottoy@kuleuven.be, liesbet.vanderperre@kuleuven.be}
}

\newcommand\blfootnote[1]{%
  \begingroup
  \renewcommand\thefootnote{}\footnote{#1}%
  \addtocounter{footnote}{-1}%
  \endgroup
}

\IEEEoverridecommandlockouts
\IEEEpubid{\makebox[\columnwidth]{%
978-1-5386-5541-2/18/\$31.00 \copyright2018 European Union
\hfill} \hspace{\columnsep}\makebox[\columnwidth]{ }
}

\maketitle

\begin{abstract}
	Both physical and MAC-layer parameters impact the autonomy of IoT devices. We present an open-source cross-layer assessment framework for Low Power Wide Area networks (LPWANs) in this paper. It extends the state-of-the-art with energy models, downlink messages, and adaptive datarate features. Hence, hypotheses and transmission schemes can be tested and evaluated. As a representative case, the LoRaWAN protocol is assessed. The findings demonstrate that a cross-layer is imperative to effectively realize LPWANs in terms of energy efficiency and throughput.
	For instance, up to a factor of three reduction in energy consumption can be achieved by transmitting longer packet on quasi-static channels. Yet, under adverse dynamic conditions, an energy penalty will occur.
\end{abstract}


\blfootnote{This paper is a preprint (IEEE ''accepted'' status). IEEE copyright notice. \copyright2019 IEEE. Personal
use of this material is permitted. Permission from IEEE must be obtained for all other uses, in
any current or future media, including reprinting/republishing this material for advertising or
promotional purposes, creating new collective works, for resale or redistribution to servers or
lists, or reuse of any copyrighted}

\blfootnote{This paper is accepted and published in the Conference Proceedings of
2019 IEEE Wireless Communications and Networking Conference (WCNC) wit doi: \url{10.1109/WCNC.2019.8885739}.}

\section{Introduction}
% !TeX root = main.tex

There is a rapidly increasing demand to interconnect devices. Sensors and actuators cooperate, often through a cloud infrastructure, enabling new applications in smart homes, cities, and sustainable environments. Many of these  Internet-of-Things (IoT) applications face strict energy constraints as they rely on battery powered devices.
Typically, a wireless IoT sensor will wake up periodically to collect measurements, e.g., temperature or chemical substances. The data can be sent immediately to the cloud infrastructure or accumulated in the node to reduce communication overhead. The device is in a low-power \textit{sleep state} for the rest of the time. Several dedicated communication systems have been developed for long range low power IoT connectivity, including LoRaWAN~\cite{semtech:lorawan_spec}, SigFox and NB-IoT~\cite{Wang:2017:PNI:3070908.3071015}.

Characterizing the energy of communication is a daunting task. It depends on numerous interrelated effects and aspects. To start with, the application determines how many bytes are in a packet and at what rate they are being sent. The latter depends, among others, on the propagation conditions. Secondly, the number of devices in the network effects the energy consumption, i.e., increasing traffic eventually will cause more collisions. A third aspect is the position of the devices and the state of the communication channel, which determines the path loss and noise. Finally, the communication protocol and radio hardware determine how long the radio is active and at what power. 
All these effects influence each other. For example, a noisy channel with a high path loss, may cause more retransmissions, which in turn increases the chance for collisions.

Simulators have been established in order to assess and optimize the  LoRaWAN protocol specifically~\cite{bor2017lora,slabicki2018adaptive, 8315103}. Here, we present a cross-layer simulation framework to realistically analyze and optimize the energy consumption of IoT devices. To this end, in extension to~\cite{bor2016lora, pop2017does}, both Adaptive Data Rate (ADR) and downlink messages have been included. We show that this has a significant impact on the number of collisions and the data extraction rate.
Secondly, the payload size and packet rate can be changed to correspond with real-life applications, whereas this was fixed in prior work~\cite{haxhibeqiri2017lora}. Hence, the framework allows application specific monitoring of the LoRaWAN Networks.
Thirdly, the energy consumption corresponding to the IoT nodes are based on measurements from a power-optimized node~\cite{lorawan-efm32}. 
Many nodes can be positioned relatively to the gateway and subjected to different signal loss and noise.
The source code (Python) of the simulator is publicly available~\cite{simulator}. We welcome researchers to tailor the simulator to their own needs.

The next section will briefly summarize the features of LoRa and LoRaWAN with an emphasis on parameters affecting the power consumption. Section~\ref{sec:simulator} highlights the key components and operation of the simulator. In Sect.~\ref{sec:results} we show the simulations and discuss the results of a case study. Conclusions and future work are presented in Sect.~\ref{sec:conclusions}.

\begin{figure*}[htbp]
	\centering
	\def\svgwidth{0.8\textwidth}
	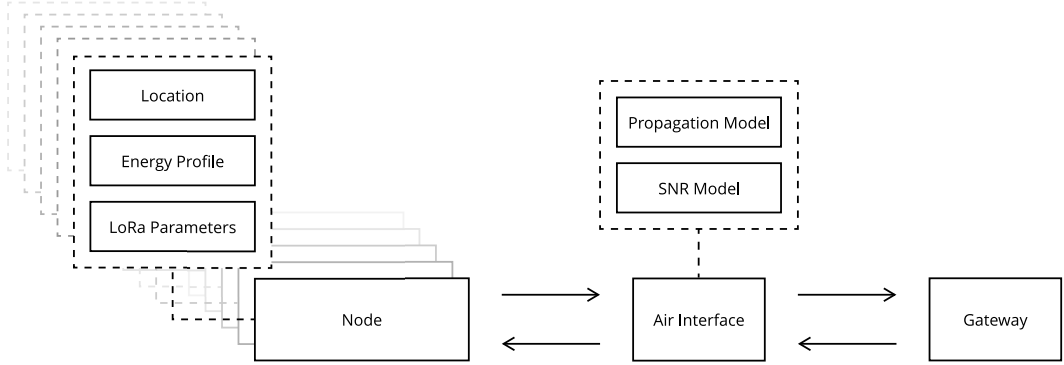
	\caption{LPWAN Simulation Framework based on a modular design for cross-layer assessment and optimization.}\label{fig:lora-simulator}
\end{figure*}

\section{LPWAN Assessment - LoRa and LoRaWAN}
% !TeX root = main.tex
%The employed modulation technique is derived from RADAR systems, where an object is located by means of sweeping a tone in frequency over time.

In our investigation, we consider the LoRa PHY and LoRaWAN MAC schemes as a first representative case. We introduce both layers with a focus on the governable parameters that affect energy consumption. Notably, a number of parameters can be adjusted in the application layer as well.

\subsubsection{LoRa PHY scheme}

LoRa, short for Long Range, is a proprietary modulation technique developed by Cycleo; later acquired by Semtech.
The modulation technique is based on Chirp Spread Spectrum (CSS), which is similar to Direct-Sequence Spread Spectrum (DSSS).
LoRa encodes information by means of chirps, in contrast to modulation with pseudorandom binary sequences in DSSS. A chirp is a sinusoidal signal whose frequency monotonically increases (\textit{upchirp}) or decreases (\textit{downchirp}).
The symbol duration is based on the spreading factor (\(SF\)) and the bandwidth (\(BW\)). Each LoRa symbol is composed of \(2^{SF}\) chirps each covering the entire bandwidth.
The symbol duration of a LoRa symbol is defined as:
\begin{equation}\label{eq:symbol-time}
T_{sym} = \dfrac{2^\text{SF}}{\text{BW}}
\end{equation}
A LoRa message consists of a preamble and data. The preamble contains only upchirps while the data part comprises upchirps with discontinuities.
The position of the discontinuities --in frequency-- is what encodes the transmitted information. 
To ensure that multiple packets can be demodulated concurrently, LoRa packets can be encoded with different orthogonal spreading factors.
This yields a robust and long-range communication link for IoT devices. 

\subsubsection{LoRaWAN MAC scheme}

On top of the physical LoRa layer, the LoRaWAN defines the multiple access control (MAC) layer and the network architecture.
Opposed to the proprietary modulation technique LoRa, LoRaWAN is an open standard specified by the LoRa Alliance.
LoRaWAN defines three device classes each targeting different use cases.
In general, LoRa devices initiate communication by means of transmitting a message to the gateway. By means of \textit{confirmed messages}, the nodes can request acknowledgments to ensure that the packets are successfully received by the gateway. % at a device-specified time instance. 
After an uplink message, the node opens two slots to receive downlink traffic from the gateway.  %These slots are only made available after an uplink message. 
This communication scheme is optimized for low power because of its uplink-centric design. LoRaWAN mandates that each LoRa device implements this scheme. The compliant devices are called class A devices.
%Class A devices are optimized for low power. It provides uplink initiated bi-directional communication. 
%LoRaWAN mandates that each LoRa device needs to support class A devices.
Class B and C devices extend the communication capabilities of class A devices by defining additional receive slots. %Both device classes have additional receive slots on top of the two receive slots subsequently to transmitting.
Class B devices have periodic receive slots while class C devices continuously listen for incoming messages. These additional downlink receive slots reduce the downlink latency yet yield a higher power consumption. %A gateway is often a Class C devices

\begin{figure*}[tbp]
	\centering
	\includegraphics[width=0.9\linewidth]{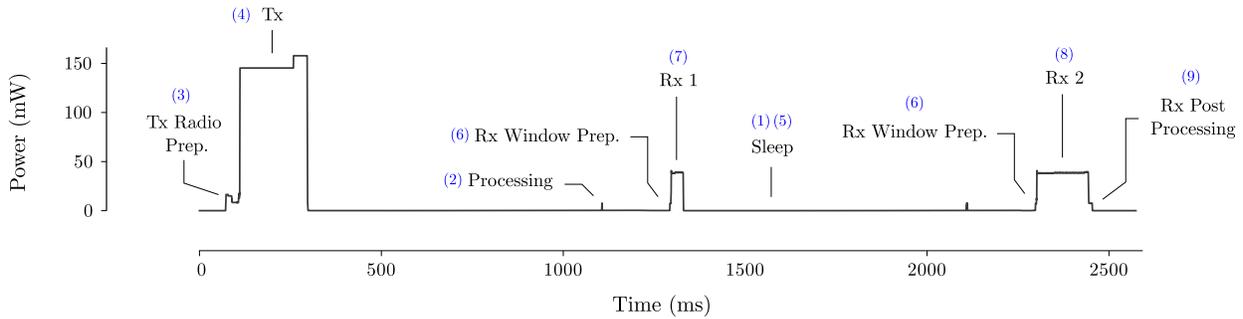}
  \caption{Measured power states of a LoRa node~\cite{lorawan-efm32}. These measurements are also summarized in Table~\ref{tab:power-states}. 
  This profile clearly shows the energy impact of transmitting a message. In this case a confirmed message was sent with SF\num{9} and a payload of \num{32} bytes.}\label{fig:energy_profile}
\end{figure*}

\subsubsection{Governable parameters}
LoRaWAN facilitates controlling the airtime~(Eq.\ref{eq:symbol-time}), data rate and energy consumption of LoRa nodes in order to optimize the overall energy consumption of the network. 
This is done by adapting the data rate and transmission power to the propagation characteristics of the LoRa link.
Increasing the spreading factor results in a higher airtime, which allows the receiver to better demodulate the message. Despite the better range, a node will consume more power when transmitting with a higher spreading factor. In addition to modifying the spreading factor, the transmission power can be altered to further increase the range or decrease the energy consumption.

%\begin{equation}\label{eq:time-on-air}
%	\begin{split}
%		T_{air} &= T_{preamble} + T_{sym} \cdot n_{payload}\\
%		T_{preamble} &= (n_{preamble} + 4.25) \cdot T_{sym} 
%	\end{split}
%\end{equation}

%The transmission power and the spreading factor of individual nodes can be improved by the adaptive data rate (ADR) mechanism. 

\begin{comment}

With \(\ceil*{\boldsymbol{\cdot}}\) ceiling, default but can be configured~\cite{semtech:lorawan_spec}, PL payload length in octets, ICRC element of 0,1 typical 1 uplink and 0 downlink~\cite{8292414}, IIH implicit header element of 0,1 and is typically set off. In the case of explicit mode, the header includes the payload length (in number of octets), the coding rate $CR$ and $PL$ and the presence of an optional 16 bit Cyclic Redundancy Check (CRC) for the payload.
copied from~\cite{8292414}: `` IDE element of
{0, 1} indicates the implementation of low data rate
optimization to increase transmission robustness to
frequency variations, which is mandated (i.e. set to 1)
for spreading factors 11 and 12 at 125 kHz and is
typically set to 0 for other spreading factors and
transmission bandwidths.''
\end{comment}

LoRaWAN devices need to comply with the regulations imposed in the industrial, scientific and medical (ISM) radio bands in which they operate. These regulations include a limitation in the duty cycle of transmissions and excited transmit power.
Concretely, LoRaWAN enforces a per band duty-cycle limitation.
After transmitting a message, the node needs to wait \(T_{\text{off}}\) seconds before transmitting again in that band as per Eq.~\ref{eq:time-off}.
Considering the case\footnote{This is a worst-case scenario, where the airtime of one packet is maximized.} of sending a message with a payload size of \num{51} bytes and a spreading factor of \num{12} and respecting a duty cycle limit of \num{1}\%, the time off is 4 minutes.
%In the worst case (SF12 with payload size 51 bytes and a duty cycle limit of 1\%), the time off is 4 minutes.

%\gilles{copied form lorawan spec}
%``The LoRaWAN enforces a per sub-band duty-cycle limitation. Each time a frame is transmitted in a given sub-band, the time of emission and the on-air duration of the frame are recorded for this sub-band. The same sub-band cannot be used again during the next Toff seconds where:''

%\gilles{fair polcy not implemented: \url{https://www.thethingsnetwork.org/docs/lorawan/duty-cycle.html}:}``Finally, on The Things Network’s public community network, we have a Fair Access Policy that limits the uplink airtime to 30 seconds per day (24 hours) per node and the downlink messages to 10 messages per day (24 hours) per node. ''

\begin{equation}\label{eq:time-off}
  T_{\text{off}} = \dfrac{T_{\text{air}}}{T_{\text{dc}}} - T_{\text{air}} \tagaddtext{[\si{\second}]}
\end{equation}
%\gilles{copied form lorawan spec}
%``During the unavailable time of a given sub-band, the device may still be able to transmit on another sub-band. If all sub-bands are unavailable, the device has to wait before any further transmission. The device adapts its channel hoping sequence according to the sub-band availability.''

\section{Cross-Layer Assessment Framework}\label{sec:simulator}
% !TeX root = main.tex

The presented cross-layer simulator (Fig.~\ref{fig:lora-simulator}) provides a generic framework to evaluate and co-optimize PHY, MAC and network parameters under realistic conditions.
To accomplish a use-case agnostic modular platform, the simulator is structured on the basis of individual components.
In the framework, each class A \textit{Node} sends LoRa packets to the \textit{Air Interface} where collision, propagation, and SNR models operate on the messages in progress.
Finally, the \textit{Gateway} receives and processes the packets.
%Depending on the simulation parameters,
%the gateway will transmit a message in the downlink to the corresponding node.
In the case of confirmed messages, the gateway will transmit a message in the downlink to the corresponding node to acknowledge the received uplink message.

\subsection{Nodes}
Each node is characterized by an energy profile, a set of LoRa parameters and a location. The default energy profile used in the simulator is based on the energy consumption of~\cite{lorawan-efm32}. We opted for this profile due to its power efficiency. Still, the simulator is not constrained to one energy profile. Different distinct profiles can be allocated to nodes, mimicking various nodes.
The different energy states of the simulated node are summarized in Table~\ref{tab:power-states} and illustrated in Figure~\ref{fig:energy_profile}.
A similar experiment has been conducted in~\cite{casals2017modeling}.
Notably, there are significant differences between the energy profile as measured on our LoRaWAN-enabled node and those reported on in~\cite{casals2017modeling}.
For instance, the node --evaluated in~\cite{casals2017modeling}-- consumes twice the power of the node in our experiments~\cite{lorawan-efm32} in transmit mode.

As can be observed from Figure~\ref{fig:energy_profile} and Table~\ref{tab:power-states}, other power states are taken into account besides transmit, receive and sleep. First, simple processing (\textit{state~2}) is simulated. Secondly, the states prior to transmitting and receiving (\textit{state~6}) are related to waking-up and setting up the radio. Finally, after receiving a downlink message, the downlink message is processed and the MAC-related functionality is executed (\textit{state~9}).

The behavior of the node is designed as specified by Semtech~\cite{semtech:modulation-basics, semtech:designers_guide, semtech:low_energy_consumption} and their LoRaWAN node implementation.\footnote{\url{http://stackforce.github.io/LoRaMac-doc/}}
In spite of the openness of the LoRaWAN MAC protocol, not all LoRa-specific documents are publicly accessible. In addition, the network operator can, to some extent, freely define the network's behavior. In our assessment, this functionality is based on the open-source implementation of The Things Network.\footnote{\url{https://github.com/thethingsnetwork}}

In order to optimize the energy budget of LoRa class A nodes, a downlink message can only be received after transmitting an uplink message. Hence, the LoRa nodes only need to listen to incoming messages at specific times. As previously mentioned, Class A LoRaWAN devices~\cite{semtech:lorawan_spec} utilize two receive windows. The data rate and center frequency of the downlink messages depends on the used receive window, the data rate and center frequency of the uplink message.
By default, the downlink message scheduled for the first receive window (RX 1) uses the same frequency and data rate as the uplink message. 
In the second receive window (RX 2), a fixed predefined frequency and data rate are being used. %These parameters are configurable through MAC commands. 
In the remainder of this paper, the receive windows will be denoted as \textit{RX 1} and \textit{RX 2}.
In the case a downlink message was received in RX 1, the node will not use the second receive slot.

In RX 2, Semtech defines a spreading factor of \num{12} while in our assessment we prefer the SF9 as proposed by The Things Network. As a lower spreading factor is favored because the base station can transmitted with higher power. The lower spreading factor results in a faster reception, which in turn yields a lower energy consumption at the node. A channel frequency of \(\SI{868.525}{\mega\hertz}\) was selected for RX 2; conform to The Things Network.
%The simulation framework utilizes the frequency and data rate specified by The Things Network, i.e. SF9 and for the second receive window. \liesbet{could we say that the simulation framework could support any mode, and that for the assessment in this paper we used the parameters from The Things Network?}

The channel frequency of the uplink packets are selected on basis of the channel availability. The end-device chooses a channel with the lowest \(T_{\text{off}}\) (Eq.~\ref{eq:time-off}). The device respects the duty cycle regulations and waits to transmit a message if the required \(T_{\text{off}}\) is not satisfied. %The devices thus hops between channels based on their availability. 
A default transmission rate (\(\lambda\)) of \num{0.02} bits per second is chosen which is equivalent to transmitting a 9 byte message every hour.

\renewcommand{\arraystretch}{1.3}

\begin{table}[]
	\renewcommand{\arraystretch}{1.3}
	\centering
	\caption{Energy profile~\cite{lorawan-efm32} used in the case.}\label{tab:power-states}
	\newcolumntype{Y}{>{\centering\arraybackslash}X}
	\begin{tabularx}{\linewidth}{@{}YYYc@{}}
		\toprule
		State & State         & Power                               & Duration                    \\
		No.   & Description   & (mW)                                & (ms)                          \\ \midrule
		1     & Sleep         & 5.7e-3                              & -                                \\
		2     & Processing    & 15                                  & 5                            \\
		3     & Tx prep.      & 12.5                                & 40                             \\
		4     & Tx            & Tab.~\ref{tab:power-consumption-tx} & Eq.~\ref{eq:symbol-time}         \\
		5a    & Wait Rx 1     & 5.7e-3                              & 1000                             \\
		5b    & Wait Rx 2     & 5.7e-3                              & 1000 - len(state 7)              \\
		6     & Rx prep.      & 8.25                                & 3.4                       \\
		7     & Rx1           & 36.96                               & airtime(DR=DR\_tx)              \\
		8     & Rx2           & 34.65                               & airtime(DR=3)                   \\
		9     & Rx post proc. & 8.3                                 & 10.7                      \\ \bottomrule
	\end{tabularx}
\end{table}

\begin{comment}
\begin{table}[tbp]
	\renewcommand{\arraystretch}{1.3}
	\centering
	\caption{Energy profile~\cite{lorawan-efm32} used in the case.}\label{tab:power-states}
	\newcolumntype{Y}{>{\centering\arraybackslash}X}
	\begin{tabularx}{\linewidth}{@{}YcYYY@{}}
		\toprule
		State & State         & Power                               & Duration                 & Energy   \\
		No.   & Description   & (mW)                                & (ms)                     & (mJ)     \\ \midrule
		1     & Sleep         & 5.7e-3                              & -                        & -        \\
		2     & Processing    & 15                                  & 5                        & 75e-3    \\
		3     & Tx prep.      & 12.5                                & 40                       & 0.5      \\
		4     & Tx            & Tab.~\ref{tab:power-consumption-tx} & Eq.~\ref{eq:symbol-time} & -        \\
		5a    & Wait Rx 1     & 5.7e-3                              & 1000                     & -        \\
		5b    & Wait Rx 2     & 5.7e-3                              & 1000 - len(state 7)      & -        \\
		6     & Rx prep.      & 8.25                                & 3.4                      & 28.05e-3 \\
		7     & Rx1           & 36.96                               & airtime( DR=DR\_tx)      & -        \\
		8     & Rx2           & 34.65                               & airtime( DR=3)           & -        \\
		9     & Rx post proc. & 8.3                                 & 10.7                     & 88.81e-3 \\ \bottomrule
	\end{tabularx}
\end{table}
\end{comment}
\begin{table}[]
	\centering
	\renewcommand{\arraystretch}{1.3}
	\caption{Measured transmit power ~\cite{lorawan-efm32} for the defined finite transmit power states.}\label{tab:power-consumption-tx}
	\begin{tabular}{@{}lccccc@{}}
		\midrule
		\textbf{Transmit Power} (\si{\deci\belmilliwatt}) & 2    & 5    & 8     & 11    & 14    \\
		\textbf{Power} (\si{\milli\watt})                 & 91.8 & 95.9 & 101.6 & 120.8 & 146.5 \\
		\bottomrule
	\end{tabular}
\end{table}

\subsection{Air Interface}
The air interface includes three main components. First, the propagation channel introduces a path loss. %and signal noise. 
Secondly, a simple SNR model is provided to translate the Received Signal Strength (RSS) to an SNR value. Finally, a collision model determines the collided packets, which occurs particularly in the uplink in a typical LPWAN case.
%The duty cycle regulations are not enforced in the air interface but on the node and the gateway.
%After processing, these messages are delivered to the gateway component. 

%tx_power_mW = {2: 91.8, 5: 95.9, 8: 101.6, 11: 120.8, 14: 146.5}

\subsubsection{Propagation Model}
Currently, the framework features two channel models.
First, a log-distance channel model with shadowing is provided, where the path loss is characterized by:
\begin{equation}\label{eq:path-loss}
	PL(d) = PL(d_0) + 10 \cdot n \log{\frac{d}{d_0}}+ X_{\sigma} \tagaddtext{[\si{\decibel}]}
\end{equation}
% gamma=2.32, d0=1000.0, std=7.8, Lpld0=128.95, GL=0
By default, the following parameters~\cite{petajajarvi2015coverage} are used:

\begin{equation}\label{eq:path-loss-defaults}
	\begin{split}
		d_0 &=  \SI{1000}{\meter} \\
		PL(d_0) &=  \SI{128.95}{\decibel} \\
		X_{\sigma} &=  \SI{7.8}{\decibel} \\
		n &= 2.32
	\end{split}
\end{equation}
An additional path loss can be included in the log-distance model to simulate indoor positioned nodes and gateways to accommodate for the additional path loss~\cite{itu-r-measurement-data} due to the penetration of a building.
Secondly, a COST 231 model~\cite{damosso1999digital} implementation can be used to model specific scenarios.

\subsubsection{SNR Model}
The current version of the simulator takes into account the noise floor, as described in~\cite{semtech:modulation-basics}. In future extensions more complex models can be included and interference could be added.
% \gilles{bijzeggen: heel moeilijk SNR te simuleren, en zeker SINR?? @Liesbet? interessante referenties of opmerkingen?} \liesbet{we can't solve world hunger - keep in mind and tackle the matter later}

\subsubsection{Collision Model}
The collision model considers the center frequency, spreading factor, timing and power to determine whether packets collide. The model is based on the findings reported in~\cite{bor2016lora}. Due to the orthogonality of the specified spreading factors, two messages encoded with different spreading factors can be demodulated concurrently without colliding.

%There a phenomenon called the capture effect was also observed for LoRa, as it is specific for frequency modulated signals. This effect translates in the demodulation of only the strongest packet if two packets with similar center frequencies collide in time. In the case of comparable signal strength, the receiver will switch between the signals, effectively unable to decode either transmissions.

\subsection{Gateway}
The gateway model is mainly based on the popular RF solution iC880A~\cite{ic880a:datasheet}.
This LoRa concentrator is able to receive up to eight packets simultaneously sent with different
spreading factors on different channels. This restriction is not considered in the assessment in this paper.
A message can be received by the gateway if it has not collided and the signal strength is higher than the sensitivity of the gateway~\cite{semtech:sx1301, ic880a:datasheet}. After demodulating the received message, the network executes Adaptive Data Rate (ADR) --if enabled-- following a mechanism inspired by
the implementation of The Things Network.\footnote{\url{https://www.thethingsnetwork.org/docs/lorawan/adr.html}}
According to the ADR specification, the network is capable of increasing the data rate and changing the transmit power of the node, while the nodes can only decrease their data rate. This can result in a low power transmit trap where nodes are no longer capable of communicating with the gateway~\cite{slabicki2018adaptive}.

Depending on the MAC LoRaWAN parameters of the uplink message,
the gateway responds with a downlink message. We currently assume that every scheduled downlink message will be received by the end-device considering gateways have a higher permitted transmit power. % \liesbet{I think we already mentioned that, and anyway people may find out if they actually would start using the framework} The downlink scheduling behavior is based on the implementation of The Things Network.\footnote{\url{https://www.thethingsnetwork.org/docs/network/architecture.html}} 
The gateway will first try to schedule a message in the receive slot which requires less energy.
%, as can be concluded from Table~\ref{tab:energy-consimption-rx-windows}. 
For instance, if a message with SF12 was sent, the gateway will try to schedule a downlink message on the second receive slot with \(SF9\) opposed to the first receive slot with \(SF12\), in order to save significant air time, and hence, energy. We measured an energy gain of four when utilizing this approach compared to using the first receive slot. This is one of the cross-layer energy optimizations already implemented in present networks.

%\gilles{as a consequenc, gateway tries to schedule messages which requires less airtime and, hence, consumes less power}

% \gilles{Copied from:\url{https://www.thethingsnetwork.org/docs/network/architecture.html}}`` In order to select the best option later, the Router additionally has to calculate a score for each option. This score is influenced by a number of factors. At the moment we consider airtime, signal strength, gateway utilization and already scheduled transmissions. ''

\section{Energy assessment and optimization: Results and Discussion}\label{sec:results}
% !TeX root = main.tex
The impact of LPWAN parameters on energy consumption and performance was assessed by performing experiments with the framework. Consequently, we analyzed the effect of the packet length on different performance parameters. Particularly, the impact of the new capabilities in our framework were assessed, most prominently the options to perform ADR and use confirmed messages.
To adequately evaluate the network, each experiment has been repeated 1000 times by means of Monto-Carlo simulations. %The authors are aware that the results will not be statistically correct with only 1000 iterations. However, our objective is to study the effects and therefore trends \gilles{todo @help wanted}.
The default parameters --for the conducted experiments-- are displayed in Table~\ref{tab:exp-defaults}.
The performance of the network and individual nodes have been evaluated based on the data extraction rate (DER), the energy per payload byte and the channel variance.

The data extraction rate defines the average ratio of the number of uniquely received packets on the base station to the uniquely transmitted packets per node. It indicates how reliable the intended payload bytes are received by the gateway.
It differs from the packet delivery success ratio because it does not include re-transmissions in the calculation of the unique transmitted packets. Hence, the DER can be improved by utilizing re-transmissions in order to accommodate for packet loss.
	      \begin{equation}
		      \mathit{DER} = \dfrac{\text{Number of uniquely received bytes}
		      }{\text{Number of uniquely transmitted bytes}}
	      \end{equation}
We prefer this parameter over the packet delivery success ratio because the main objective is that the intended payload is received by the gateway. The number of re-transmissions necessary to achieve this goal is of secondary importance. 

In the experiments, the energy per payload byte indicates the impact of the set of defined parameters on energy efficiency. 
	
The channel is characterized by its variance \(\sigma\) (in \(\si{\decibel}\)) according to Eq.~\ref{eq:path-loss}. This variance with respect to the average path loss is a consequence of varying propagation characteristics in both space and time. In the following experiments, the defaults of Eq.~\ref{eq:path-loss-defaults} are used; if not stated otherwise.

%\gilles{wat allemaal bekeken, impact van ADR/CONF/payload size/channel variance/ ADR-NET algortihm. Of is het genoeg om gewoon de vorige pargrafen te lezen zonder deze uiteenzetting?}

\begin{table}[hb]
	\renewcommand{\arraystretch}{1.3}
	\centering
	\caption{Parameters for the experiments reported on.}\label{tab:exp-defaults}
	\begin{tabular}{@{}ll@{}}
		\toprule
		Parameter                          & Default Value                                        \\ \midrule
		Channel Variance \(\sigma\)        & \SI{7.8}{\decibel}                                   \\
		Number of Nodes                    & \num{100}                                            \\
		Data Transmission rate \(\lambda\) & \SI{0.02}{bps}                                       \\
		Initial Transmit Power             & \SI{14}{\dBm}                                        \\
		Channels                           & \num{868.1}, \num{868.3} and \SI{868.5}{\mega\hertz} \\
		RX2 Channel                        & \SI{868.525}{\mega\hertz}                            \\
		RX2 Data Rate                      & DR3 (SF 9)                                           \\
		Cell Radius                        & \SI{1000}{\meter}                                    \\ \bottomrule
	\end{tabular}
\end{table}

% =======================================================
% ======================= FIGURES =======================
% =======================================================
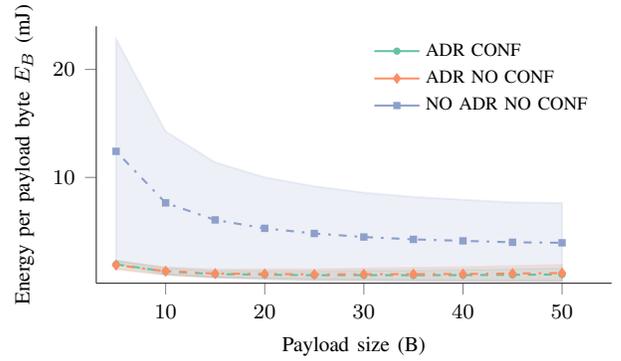
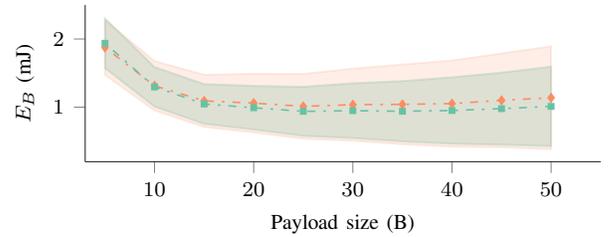
\begin{figure}[tbp]
	\centering
	\begin{subfigure}[b]{0.5\textwidth}
		\centering
		% !TeX root = main.tex

% This file was created by matplotlib2tikz v0.6.15.
\begin{tikzpicture}

\definecolor{color0}{rgb}{0.4,0.76078431372549,0.647058823529412}
\definecolor{color1}{rgb}{0.988235294117647,0.552941176470588,0.384313725490196}
\definecolor{color2}{rgb}{0.552941176470588,0.627450980392157,0.796078431372549}
\definecolor{color3}{rgb}{0.905882352941176,0.541176470588235,0.764705882352941}

\begin{axis}[
xlabel={Payload size (B)},
ylabel={Energy per payload byte $E_B$ (mJ)},
xmin=3, xmax=55,
ymin=0.2, ymax=24,
height=5cm,
width=0.93\textwidth,
tick align=outside,
tick pos=left,
x grid style={white!80.0!black},
y grid style={white!80.0!black},
axis line style={white!15.0!black},
axis x line*={bottom},
axis y line*={left},
legend cell align={left},
legend entries={{ADR CONF},{ADR NO CONF},{NO ADR NO CONF}},
legend style={draw=none, font=\scriptsize}
]
\addlegendimage{mark=*, mark size=1,color0}
\addlegendimage{mark=diamond*, mark size=1.5, color1}
\addlegendimage{mark=square*, mark size=1,color2}
\path [draw=color0, fill=color0, opacity=0.15] (axis cs:5,2.30299185672114)
--(axis cs:5,1.57166852718731)
--(axis cs:10,1.00734584486536)
--(axis cs:15,0.754534637392449)
--(axis cs:20,0.670989483568726)
--(axis cs:25,0.578984625360715)
--(axis cs:30,0.546256966152437)
--(axis cs:35,0.497069202854129)
--(axis cs:40,0.464059464268145)
--(axis cs:45,0.449361138923774)
--(axis cs:50,0.428546146626067)
--(axis cs:50,1.59482422702312)
--(axis cs:50,1.59482422702312)
--(axis cs:45,1.50733117716424)
--(axis cs:40,1.4387084476162)
--(axis cs:35,1.3831914045942)
--(axis cs:30,1.35104095817237)
--(axis cs:25,1.29928591262843)
--(axis cs:20,1.3124886501429)
--(axis cs:15,1.33918180132729)
--(axis cs:10,1.58807428691235)
--(axis cs:5,2.30299185672114)
--cycle;

\path [draw=color1, fill=color1, opacity=0.15] (axis cs:5,2.27701967513883)
--(axis cs:5,1.47367453075879)
--(axis cs:10,0.950882819984792)
--(axis cs:15,0.709270995975777)
--(axis cs:20,0.628199434469574)
--(axis cs:25,0.537531547717974)
--(axis cs:30,0.507726115076829)
--(axis cs:35,0.456006916254134)
--(axis cs:40,0.420344595406092)
--(axis cs:45,0.412162875936439)
--(axis cs:50,0.382555861806964)
--(axis cs:50,1.8931871829514)
--(axis cs:50,1.8931871829514)
--(axis cs:45,1.78919190891175)
--(axis cs:40,1.6850484970174)
--(axis cs:35,1.62698200003252)
--(axis cs:30,1.56601028134288)
--(axis cs:25,1.48751085056689)
--(axis cs:20,1.49130531416935)
--(axis cs:15,1.47572748756964)
--(axis cs:10,1.67704344888741)
--(axis cs:5,2.27701967513883)
--cycle;

\path [draw=color2, fill=color2, opacity=0.15] (axis cs:5,22.8159209470032)
--(axis cs:5,2.01056391798319)
--(axis cs:10,1.0311772268272)
--(axis cs:15,0.72859400110202)
--(axis cs:20,0.56898721762832)
--(axis cs:25,0.434764392076499)
--(axis cs:30,0.369900554268213)
--(axis cs:35,0.327540581730993)
--(axis cs:40,0.315947031486924)
--(axis cs:45,0.29223308450319)
--(axis cs:50,0.26052877434271)
--(axis cs:50,7.62700598656362)
--(axis cs:50,7.62700598656362)
--(axis cs:45,7.69290171197119)
--(axis cs:40,7.93441973555564)
--(axis cs:35,8.19300214118666)
--(axis cs:30,8.5906092402071)
--(axis cs:25,9.17957470939652)
--(axis cs:20,10.0010172490506)
--(axis cs:15,11.3954718567872)
--(axis cs:10,14.2557902935208)
--(axis cs:5,22.8159209470032)
--cycle;

\addplot [thick, color0, dashed, mark=*, mark size=1, mark options={solid}]
table {%
5 1.93733019195423
10 1.29771006588885
15 1.04685821935987
20 0.991739066855813
25 0.93913526899457
30 0.948648962162402
35 0.940130303724163
40 0.951383955942173
45 0.978346158044005
50 1.0116851868246
};
\addplot [thick, color1, dash pattern=on 1pt off 3pt on 3pt off 3pt, mark=diamond*, mark size=1.5, mark options={solid}]
table {%
5 1.87534710294881
10 1.3139631344361
15 1.09249924177271
20 1.05975237431946
25 1.01252119914243
30 1.03686819820985
35 1.04149445814332
40 1.05269654621175
45 1.1006773924241
50 1.13787152237918
};
\addplot [thick, color2, dash pattern=on 1pt off 3pt on 3pt off 3pt, mark=square*, mark size=1, mark options={solid}]
table {%
5 12.4132424324932
10 7.643483760174
15 6.06203292894459
20 5.28500223333947
25 4.80716955073651
30 4.48025489723766
35 4.26027136145883
40 4.12518338352128
45 3.99256739823719
50 3.94376738045317
};
\end{axis}

\end{tikzpicture}
		\caption{Energy consumption per transmitted payload byte for different configurations. Disabling adaptive data rate may result in up to an order of magnitude higher average energy consumption and considerably increase its spread.}\label{fig:payload_vs_energy_a}
		%\vspace*{5mm} % add a little space
	\end{subfigure}%
	\\[3ex]
	\begin{subfigure}[b]{0.5\textwidth}
		\centering
		% This file was created by matplotlib2tikz v0.6.15.
\begin{tikzpicture}

\definecolor{color0}{rgb}{0.4,0.76078431372549,0.647058823529412}
\definecolor{color1}{rgb}{0.988235294117647,0.552941176470588,0.384313725490196}

\begin{axis}[
    xlabel={Payload size (B)},
    ylabel={$E_B$ (mJ)},
    y=\textwidth/10,
    xmin=3, xmax=55,
    ymin=0.2, ymax=2.5,
    tick align=outside,
    tick pos=left,
    x grid style={white!80.0!black},
    y grid style={white!80.0!black},
    axis line style={white!15.0!black},
    axis x line*={bottom},
    axis y line*={left}
]
\addlegendimage{mark=*, dashed,color1}
\addlegendimage{mark=*, dashed,color0}

\path [draw=color0, fill=color0, opacity=0.25] (axis cs:5,2.30299185672114)
--(axis cs:5,1.57166852718731)
--(axis cs:10,1.00734584486536)
--(axis cs:15,0.754534637392449)
--(axis cs:20,0.670989483568726)
--(axis cs:25,0.578984625360715)
--(axis cs:30,0.546256966152437)
--(axis cs:35,0.497069202854129)
--(axis cs:40,0.464059464268145)
--(axis cs:45,0.449361138923774)
--(axis cs:50,0.428546146626067)
--(axis cs:50,1.59482422702312)
--(axis cs:50,1.59482422702312)
--(axis cs:45,1.50733117716424)
--(axis cs:40,1.4387084476162)
--(axis cs:35,1.3831914045942)
--(axis cs:30,1.35104095817237)
--(axis cs:25,1.29928591262843)
--(axis cs:20,1.3124886501429)
--(axis cs:15,1.33918180132729)
--(axis cs:10,1.58807428691235)
--(axis cs:5,2.30299185672114)
--cycle;

\path [draw=color1, fill=color1, opacity=0.15] (axis cs:5,2.27701967513883)
--(axis cs:5,1.47367453075879)
--(axis cs:10,0.950882819984792)
--(axis cs:15,0.709270995975777)
--(axis cs:20,0.628199434469574)
--(axis cs:25,0.537531547717974)
--(axis cs:30,0.507726115076829)
--(axis cs:35,0.456006916254134)
--(axis cs:40,0.420344595406092)
--(axis cs:45,0.412162875936439)
--(axis cs:50,0.382555861806964)
--(axis cs:50,1.8931871829514)
--(axis cs:50,1.8931871829514)
--(axis cs:45,1.78919190891175)
--(axis cs:40,1.6850484970174)
--(axis cs:35,1.62698200003252)
--(axis cs:30,1.56601028134288)
--(axis cs:25,1.48751085056689)
--(axis cs:20,1.49130531416935)
--(axis cs:15,1.47572748756964)
--(axis cs:10,1.67704344888741)
--(axis cs:5,2.27701967513883)
--cycle;

\addplot [semithick, color1, dash pattern=on 1pt off 3pt on 3pt off 3pt, mark=diamond*, mark size=1.5, mark options={solid}]
table {%
5 1.87534710294881
10 1.3139631344361
15 1.09249924177271
20 1.05975237431946
25 1.01252119914243
30 1.03686819820985
35 1.04149445814332
40 1.05269654621175
45 1.1006773924241
50 1.13787152237918
};

\addplot [semithick, dash pattern=on 1pt off 3pt on 3pt off 3pt, color0, mark=square*, mark size=1, mark options={solid}]
table {%
5 1.93733019195423
10 1.29771006588885
15 1.04685821935987
20 0.991739066855813
25 0.93913526899457
30 0.948648962162402
35 0.940130303724163
40 0.951383955942173
45 0.978346158044005
50 1.0116851868246
};
\end{axis}

\end{tikzpicture}
		\caption{Zoom in showing the further energy reduction by disabling confirmed messages, especially reducing the spread for larger packet sizes. }\label{fig:payload_vs_energy_b}
	\end{subfigure}
	\caption{Energy per transmitted payload byte as a function of payload size. The average is depicted by the markers while the deviation is illustrated by the shaded area. The experiment simulated 30 operation days.}\label{fig:payload_vs_energy}
\end{figure}

\begin{comment}
\begin{figure}[]
	\centering
	\input{Figures/collisions_per_payload.tex}
	\caption{Collisions ratio (w.r.t. the total bytes sent) for different configurations of ADR and confirmed messages. }\label{fig:collisions_vs_adr_conf}
\end{figure}

\begin{figure}[h]
	\centering
	\begin{subfigure}[b]{0.5\textwidth}
		\centering
		\input{Figures/impact_snr_max_min_a.tex}
		\caption{The effect of utilizing a minimum SNR selection criteria instead of the current maximum criteria on the DER is expressed as: $\alpha = DER_{max, SNR}$ /  $DER_{min, SNR}$. A value higher than 100\% indicates that the maximum selection performs better than the minimum selection in terms of the data extraction rate.}
	\end{subfigure}%
	\\[3ex]
	\begin{subfigure}[b]{0.5\textwidth}
		\centering
		\input{Figures/impact_snr_max_min_b.tex}
		\caption{$\beta = E_{B,max, SNR}$ /  $E_{B,min, SNR}$ gives a measure to compare the energy efficiency between the minumum and maximum selection criteria.}
	\end{subfigure}
	\caption{Ratio between the maximum and the minimum SNR selection criteria in ADR for the (a) data extraction rate and (b) energy. The \(100 \%\) line is a baseline for comparison indicating which technique ought to be selected. %(20 simulations each 30 simulated days long)
	}\label{fig:impact-snr-max-min}
\end{figure}

\end{comment}

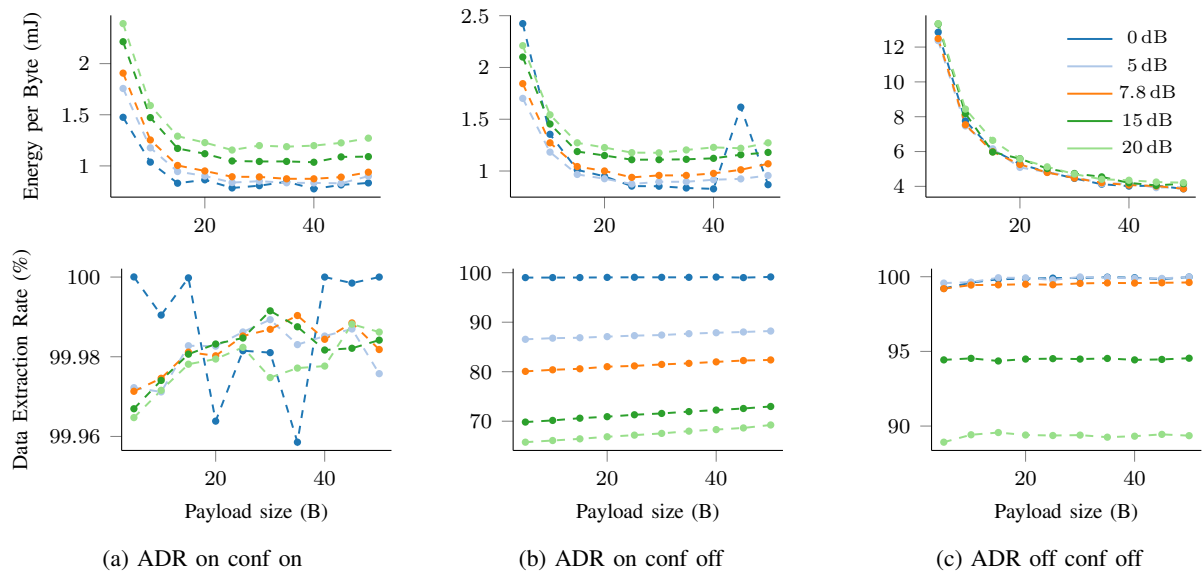
\begin{figure*}[tbp]
	\centering
	\begin{subfigure}[b]{0.3\linewidth}
		\centering
		% This file was created by matplotlib2tikz v0.6.15.
\begin{tikzpicture}

\definecolor{color0}{rgb}{0.12156862745098,0.466666666666667,0.705882352941177}
\definecolor{color1}{rgb}{0.682352941176471,0.780392156862745,0.909803921568627}
\definecolor{color2}{rgb}{1,0.498039215686275,0.0549019607843137}
\definecolor{color3}{rgb}{1,0.733333333333333,0.470588235294118}
\definecolor{color4}{rgb}{0.172549019607843,0.627450980392157,0.172549019607843}
\definecolor{color5}{rgb}{0.596078431372549,0.874509803921569,0.541176470588235}

\begin{axis}[
ylabel={Energy per Byte (mJ)},
xmin=2.75, xmax=52.25,
ymin=0.694768797453648, ymax=2.47197893240613,
tick align=outside,
tick pos=left,
x grid style={white!69.01960784313725!black},
y grid style={white!69.01960784313725!black},
height=4cm,
width=0.95\textwidth
]
\addplot [color0, dashed, mark=*,   mark options={solid}, forget plot]
table {%
5 1.4750884458488
10 1.03682179804677
15 0.83080068756251
20 0.863786313993537
25 0.784503368681476
30 0.806692736488687
35 0.845055728030364
40 0.775551076315125
45 0.814078649565028
50 0.833557815220113
};
\addplot [color1, dashed, mark=*,   mark options={solid}, forget plot]
table {%
5 1.7570082636473
10 1.17688233625459
15 0.945057015764643
20 0.901571290658206
25 0.837188018982421
30 0.850766302851482
35 0.834976698008423
40 0.833475829252397
45 0.829884725110207
50 0.899752079662383
};
\addplot [color2, dashed, mark=*,   mark options={solid}, forget plot]
table {%
5 1.90711158615702
10 1.25351895072853
15 1.00442360758012
20 0.950610028768001
25 0.893888241833968
30 0.892488998873549
35 0.873948715650174
40 0.873897940154721
45 0.890407840890781
50 0.938608201653661
};

\addplot [color4, dashed, mark=*,   mark options={solid}, forget plot]
table {%
5 2.21425505493905
10 1.47229889526904
15 1.17074401671981
20 1.11911746100203
25 1.04803388174811
30 1.04334479525228
35 1.043790980887
40 1.03592257762454
45 1.08803616946763
50 1.0906149821655
};
\addplot [color5, dashed, mark=*,   mark options={solid}, forget plot]
table {%
5 2.39119665354465
10 1.59042708053051
15 1.29060229294306
20 1.22771781161326
25 1.15518588965675
30 1.19862383770603
35 1.18812659915983
40 1.19826564935025
45 1.22533016581613
50 1.27126775820504
};
\end{axis}

\end{tikzpicture}
	\end{subfigure}
	\begin{subfigure}[b]{0.3\linewidth}
		\centering
		% This file was created by matplotlib2tikz v0.6.15.
\begin{tikzpicture}

\definecolor{color0}{rgb}{0.12156862745098,0.466666666666667,0.705882352941177}
\definecolor{color1}{rgb}{0.682352941176471,0.780392156862745,0.909803921568627}
\definecolor{color2}{rgb}{1,0.498039215686275,0.0549019607843137}
\definecolor{color3}{rgb}{1,0.733333333333333,0.470588235294118}
\definecolor{color4}{rgb}{0.172549019607843,0.627450980392157,0.172549019607843}
\definecolor{color5}{rgb}{0.596078431372549,0.874509803921569,0.541176470588235}

\begin{axis}[
xmin=2.75, xmax=52.25,
ymin=0.747803661840947, ymax=2.50329042352313,
tick align=outside,
tick pos=left,
x grid style={white!69.01960784313725!black},
y grid style={white!69.01960784313725!black},
height=4cm,
width=0.95\textwidth
]
\addplot [color0, dashed, mark=*,   mark options={solid}, forget plot]
table {%
5 2.4234955707194
10 1.35475421830674
15 1.01107445993034
20 0.948286884736364
25 0.854260439848536
30 0.85306317323071
35 0.836003352660744
40 0.827598514644683
45 1.61678977794279
50 0.867981446020637
};
\addplot [color1, dashed, mark=*,   mark options={solid}, forget plot]
table {%
5 1.70130399698612
10 1.18255125505788
15 0.96693690619834
20 0.925665059602878
25 0.874607982274425
30 0.895641105571106
35 0.896293063871687
40 0.91589679273204
45 0.923971307916621
50 0.95630678428046
};
\addplot [color2, dashed, mark=*,   mark options={solid}, forget plot]
table {%
5 1.84366944114251
10 1.27222510219254
15 1.04311573298757
20 1.00042914160454
25 0.938408684888022
30 0.957761718097454
35 0.957485594012515
40 0.976794063228471
45 1.01171176981934
50 1.06978354844107
};

\addplot [color4, dashed, mark=*,   mark options={solid}, forget plot]
table {%
5 2.10052639291128
10 1.45384206879301
15 1.18815654458605
20 1.14972000754574
25 1.10925522871482
30 1.10979895204477
35 1.11300850600871
40 1.1225155939668
45 1.15726033883013
50 1.17941298801994
};
\addplot [color5, dashed, mark=*,   mark options={solid}, forget plot]
table {%
5 2.21054237076718
10 1.54282750789802
15 1.27237004094093
20 1.22648230432076
25 1.17757209764276
30 1.17603653526423
35 1.20342034182802
40 1.22817001131003
45 1.21968814533584
50 1.27237443813148
};
\end{axis}

\end{tikzpicture}
	\end{subfigure}
	\begin{subfigure}[b]{0.3\linewidth}
		\centering
		% This file was created by matplotlib2tikz v0.6.15.
\begin{tikzpicture}

\definecolor{color0}{rgb}{0.12156862745098,0.466666666666667,0.705882352941177}
\definecolor{color1}{rgb}{0.682352941176471,0.780392156862745,0.909803921568627}
\definecolor{color2}{rgb}{1,0.498039215686275,0.0549019607843137}
\definecolor{color3}{rgb}{1,0.733333333333333,0.470588235294118}
\definecolor{color4}{rgb}{0.172549019607843,0.627450980392157,0.172549019607843}
\definecolor{color5}{rgb}{0.596078431372549,0.874509803921569,0.541176470588235}

\begin{axis}[
xmin=2.75, xmax=52.25,
ymin=3.38129176318195, ymax=13.8216318473626,
tick align=outside,
tick pos=left,
x grid style={white!69.01960784313725!black},
y grid style={white!69.01960784313725!black},
height=4cm,
width=0.95\textwidth,
legend entries={
{$\SI{0}{\decibel}$},
{$\SI{5}{\decibel}$},
{$\SI{7.8}{\decibel}$},
{$\SI{15}{\decibel}$},
{$\SI{20}{\decibel}$}
},
legend columns=1
]
\addlegendimage{color0}
\addlegendimage{color1}
\addlegendimage{color2}
%\addlegendimage{color3}
\addlegendimage{color4}
\addlegendimage{color5}
\addplot [color0, dashed, mark=*,   mark options={solid}, forget plot]
table {%
5 12.8514311993824
10 7.75757445073722
15 6.09348292263053
20 5.2777252072885
25 4.82715338209399
30 4.45631595644929
35 4.12123198361873
40 4.01433083973713
45 4.04064352646082
50 3.85585267609925
};
\addplot [color1, dashed, mark=*,   mark options={solid}, forget plot]
table {%
5 12.3700178827088
10 7.47062609781687
15 6.15572246364926
20 5.08725694642894
25 4.81368439015104
30 4.55691301121195
35 4.15543831708373
40 4.16819322229756
45 3.92564714798751
50 3.95640568062556
};
\addplot [color2, dashed, mark=*,   mark options={solid}, forget plot]
table {%
5 12.4862701278103
10 7.53894341400922
15 5.98809362805015
20 5.25739946509246
25 4.79671457389751
30 4.47402098015479
35 4.20937106760853
40 4.07282486541202
45 4.01665573282113
50 3.86333033642702
};

\addplot [color4, dashed, mark=*,   mark options={solid}, forget plot]
table {%
5 13.3245308243138
10 8.18273200334257
15 5.97010312777708
20 5.58821167516522
25 5.02364989528775
30 4.73872287460286
35 4.54587686296606
40 4.19958192555314
45 4.05428626339613
50 4.15359923426753
};
\addplot [color5, dashed, mark=*,   mark options={solid}]
table {%
5 13.3470709344453
10 8.4358784278641
15 6.64188839714607
20 5.58016976030663
25 5.12143705025077
30 4.70557145694408
35 4.40634242070086
40 4.34966845458284
45 4.2488564781956
50 4.2108458208394
};
\end{axis}
\end{tikzpicture}
	\end{subfigure}\\
	\begin{subfigure}[b]{0.3\linewidth}
		\centering
		% This file was created by matplotlib2tikz v0.6.15.
\begin{tikzpicture}

\definecolor{color0}{rgb}{0.12156862745098,0.466666666666667,0.705882352941177}
\definecolor{color1}{rgb}{0.682352941176471,0.780392156862745,0.909803921568627}
\definecolor{color2}{rgb}{1,0.498039215686275,0.0549019607843137}
\definecolor{color3}{rgb}{1,0.733333333333333,0.470588235294118}
\definecolor{color4}{rgb}{0.172549019607843,0.627450980392157,0.172549019607843}
\definecolor{color5}{rgb}{0.596078431372549,0.874509803921569,0.541176470588235}

\begin{axis}[
xlabel={Payload size (B)},
ylabel={Data Extraction Rate (\%)},
xmin=2.75, xmax=52.25,
ymin=99.9564884275254, ymax=100.002107186579,
tick align=outside,
tick pos=left,
x grid style={white!69.01960784313725!black},
y grid style={white!69.01960784313725!black},
height=4cm,
width=0.95\textwidth
]
\addplot [color0, dashed, mark=*,   mark options={solid}, forget plot]
table {%
5 100.000033606622
10 99.9904615020851
15 99.9997984496749
20 99.9638492365335
25 99.981515587401
30 99.9810354254218
35 99.9585620074824
40 100
45 99.9984858854239
50 100
};
\addplot [color1, dashed, mark=*,   mark options={solid}, forget plot]
table {%
5 99.9722393159176
10 99.9711782454202
15 99.9827668393678
20 99.9826605911529
25 99.9862177461645
30 99.9893084590262
35 99.9830453395877
40 99.9851946366898
45 99.9869722630391
50 99.9757579838723
};
\addplot [color2, dashed, mark=*,   mark options={solid}, forget plot]
table {%
5 99.9713320770532
10 99.9745951010215
15 99.9811531263134
20 99.9802422541572
25 99.9852138364886
30 99.9868846901962
35 99.9903471267398
40 99.984381900238
45 99.9884924852901
50 99.9818168961442
};

\addplot [color4, dashed, mark=*,   mark options={solid}, forget plot]
table {%
5 99.966952251214
10 99.9740631488412
15 99.9806491797066
20 99.9831972986633
25 99.9847065506381
30 99.9915246883809
35 99.9875208966118
40 99.9816937088277
45 99.9821280900987
50 99.9841723129976
};
\addplot [color5, dashed, mark=*,   mark options={solid}, forget plot]
table {%
5 99.9647293236167
10 99.9715820914774
15 99.978124911794
20 99.9794330235741
25 99.9823543053814
30 99.974779878782
35 99.9771570675327
40 99.9776611492397
45 99.9881830356872
50 99.9861953326757
};
\end{axis}

\end{tikzpicture}
		\caption{ADR on conf on}
	\end{subfigure}
	\begin{subfigure}[b]{0.3\linewidth}
		\centering
		% This file was created by matplotlib2tikz v0.6.15.
\begin{tikzpicture}

\definecolor{color0}{rgb}{0.12156862745098,0.466666666666667,0.705882352941177}
\definecolor{color1}{rgb}{0.682352941176471,0.780392156862745,0.909803921568627}
\definecolor{color2}{rgb}{1,0.498039215686275,0.0549019607843137}
\definecolor{color3}{rgb}{1,0.733333333333333,0.470588235294118}
\definecolor{color4}{rgb}{0.172549019607843,0.627450980392157,0.172549019607843}
\definecolor{color5}{rgb}{0.596078431372549,0.874509803921569,0.541176470588235}

\begin{axis}[
xlabel={Payload size (B)},
xmin=2.75, xmax=52.25,
ymin=64.0943885473461, ymax=100.814522782605,
tick align=outside,
tick pos=left,
x grid style={white!69.01960784313725!black},
y grid style={white!69.01960784313725!black},
height=4cm,
width=0.95\textwidth
]
\addplot [color0, dashed, mark=*,   mark options={solid}, forget plot]
table {%
5 99.0112368459157
10 99.0142451750214
15 99.0192490520651
20 99.0519243797787
25 99.0723300293877
30 99.0630541931553
35 99.0651785483078
40 99.1157184081316
45 99.0072576857225
50 99.1454257719115
};
\addplot [color1, dashed, mark=*,   mark options={solid}, forget plot]
table {%
5 86.5438688077852
10 86.7765480409116
15 86.8601428902526
20 87.0853951827669
25 87.284383165713
30 87.4048057583564
35 87.6819033804595
40 87.8728837961555
45 88.0238831651556
50 88.2151584304381
};
\addplot [color2, dashed, mark=*,   mark options={solid}, forget plot]
table {%
5 80.0662678009452
10 80.3818625692454
15 80.5888204943157
20 81.015222013456
25 81.1711065212167
30 81.4722230628777
35 81.6883563418161
40 81.983211280039
45 82.2564863759599
50 82.3769097882859
};

\addplot [color4, dashed, mark=*,   mark options={solid}, forget plot]
table {%
5 69.8253421527942
10 70.1581290592215
15 70.6072435414796
20 70.9141523591744
25 71.3006218092143
30 71.5701011894553
35 71.9419190729662
40 72.2510369255863
45 72.5682106647037
50 72.9759303464832
};
\addplot [color5, dashed, mark=*,   mark options={solid}, forget plot]
table {%
5 65.7634855580397
10 66.0989815660773
15 66.438314054218
20 66.8566966763563
25 67.188340324836
30 67.5385319824027
35 67.9925125379671
40 68.3031676644128
45 68.6385334884172
50 69.2369331878847
};
\end{axis}

\end{tikzpicture}
		\caption{ADR on conf off}
	\end{subfigure}
	\begin{subfigure}[b]{0.3\linewidth}
		\centering
		% This file was created by matplotlib2tikz v0.6.15.
\begin{tikzpicture}

\definecolor{color0}{rgb}{0.12156862745098,0.466666666666667,0.705882352941177}
\definecolor{color1}{rgb}{0.682352941176471,0.780392156862745,0.909803921568627}
\definecolor{color2}{rgb}{1,0.498039215686275,0.0549019607843137}
\definecolor{color3}{rgb}{1,0.733333333333333,0.470588235294118}
\definecolor{color4}{rgb}{0.172549019607843,0.627450980392157,0.172549019607843}
\definecolor{color5}{rgb}{0.596078431372549,0.874509803921569,0.541176470588235}

\begin{axis}[
xlabel={Payload size (B)},
xmin=2.75, xmax=52.25,
ymin=88.3675401663808, ymax=100.539113140735,
tick align=outside,
tick pos=left,
x grid style={white!69.01960784313725!black},
y grid style={white!69.01960784313725!black},
height=4cm,
width=0.95\textwidth
]
\addplot [color0, dashed, mark=*,   mark options={solid}, forget plot]
table {%
5 99.2143350248075
10 99.6056433061177
15 99.829294959969
20 99.8698846032354
25 99.9073913781541
30 99.9156778652379
35 99.971273772719
40 99.9348337722629
45 99.8370383383968
50 99.9811465585736
};
\addplot [color1, dashed, mark=*,   mark options={solid}, forget plot]
table {%
5 99.5744088900358
10 99.6378437494962
15 99.9306370503002
20 99.9272959886469
25 99.7859979087305
30 99.9840573451242
35 99.9451285710585
40 99.8915293288007
45 99.8860937424265
50 99.9858598237191
};
\addplot [color2, dashed, mark=*,   mark options={solid}, forget plot]
table {%
5 99.2002807235488
10 99.4407663268879
15 99.4555829454958
20 99.4979845457197
25 99.4620222317086
30 99.5531210910705
35 99.5814904870293
40 99.5743409149651
45 99.5934910095478
50 99.6150828262282
};

\addplot [color4, dashed, mark=*,   mark options={solid}, forget plot]
table {%
5 94.4313285107916
10 94.5334297402925
15 94.3578038534087
20 94.4867633122971
25 94.5138636760929
30 94.4862347557285
35 94.527704174763
40 94.4318974080126
45 94.4620818767305
50 94.5422529283625
};
\addplot [color5, dashed, mark=*,   mark options={solid}, forget plot]
table {%
5 88.9207934833969
10 89.4166545165753
15 89.5661115367555
20 89.4006057257619
25 89.367384325787
30 89.3940219090438
35 89.2557759390244
40 89.3214497280414
45 89.4435594749944
50 89.3565148285465
};
\end{axis}

\end{tikzpicture}
		\caption{ADR off conf off}
	\end{subfigure}
	\caption{Impact of the channel variance and payload size on the energy and data extraction rate.}\label{fig:impact-adr-conf-variance}
\end{figure*}

% =======================================================
% ======================= ./FIGURES =====================
% =======================================================

%\geof{
%  ik denk dat het in de conclusies belangrijk is van de minder evidente %findings nog eens in de verf te zetten. Er zijn er namelijk een aantal die %een bevestiging zijn van wat al was aangetoond. Misschien moeten we die %trouwens eerst zetten. Ter bevestiging van "de simulator werkt", maar doordat %de onze beter is, hebben we ook nog een aantal nieuwe findings.
%}

\subsection{Validation of the Simulation Model}
The cross-layer simulation framework has been evaluated and validated by checking the results with findings from related work.
As advised by Semtech and The Things Network, the ADR rate should only be enabled when a node has a fixed location. This is confirmed by our simulations.
If the channel is dynamic, the effect of ADR is nullified and even reduces the data extraction rate (Fig.~\ref{fig:impact-adr-conf-variance}), i.e.\ fewer packets are successfully received by the gateway. %\gilles{figuur lijkt mij hier overbodig vermits het gaat over de validatie van het framework.}
Furthermore, the impact of the duty cycle limit on the downlink capabilities of the gateway has been assessed as well. The experiments confirm the findings reported in~\cite{pop2017does}.
If only the default channels (Table~\ref{tab:exp-defaults}) are utilized, the gateway is incapable of acknowledging all confirmed messages. Consequently, the number of retransmitted packages increases, which in its turn yields a lower DER, as also observed in~\cite{pop2017does}. Hence, the scalability of the network is mainly constrained by (I) employing confirmed messages and (II) the duty cycle limitation.

\subsection{Results - Cross-Layer Approach to the Rescue}

% \liesbet{I very much like the findings-matter yet would keep that as the core of another paper}\textbf{Finding \#1 - Nodes sending smaller packets are more power hungry}\\
We have assessed the impact of package length as a first cross-layer optimization opportunity. As expected, the average energy consumption per payload byte decreases when sending larger packets (Fig.~\ref{fig:payload_vs_energy}). To save energy, non-time-critical data can be accumulated, because by increasing the payload size
\begin{enumerate}
	\item the overhead related to header information decreases,
	\item the overhead of starting and initializing a transmission lowers,
	\item the number of retransmissions in a stable propagation environment reduces,
	\item the number of downlink receive windows is also lower.
\end{enumerate}

Substantial energy savings, up to an order of magnitude, can be achieved by enabling ADR as indicated in Fig.~\ref{fig:payload_vs_energy_a}. This obviously demonstrates the importance of including ADR in the assessment and optimization of transmission parameters in LPWANs to ensure long battery lifetime of IoT nodes.

%\textbf{Finding \#2 - Increasing the payload size results in more bytes which are sent sub-optimal}\\
Despite the aforementioned beneficial effects of increasing the payload size, %the downside of 
sending more bytes per packet increases the total number of bytes which are sent sub-optimal.
 %with sub-optimal parameters when increasing the payload size should be considered.
Only after receiving 20 uplink messages, the network will respond with the adequate ADR parameters to accommodate for non-optimal propagation matched LoRa parameters.
For higher payload sizes this implies that more bytes have been sent before the LoRa parameters are adjusted to the channel. In addition, ADR changes the parameters in steps yielding an even slower adaption to the propagation environment for larger payload sizes. This effect is clearly notable when observing the energy consumption over a short time period or when nodes have a slow data transmission rate. The phenomenon results in a higher energy spread as depicted in Figure~\ref{fig:payload_vs_energy_b}. In quasi-static situations the impact will become negligible on the longer term. In dynamic situations, however, the trade-off on packet length may yield a different result.

To faster adapt to the channel, LoRa devices could first sent 20 smaller packets. % at a higher rate (close to the duty cycle limit)
This will result in reduced airtime and energy for packets which are sent with non-optimal parameters.
The further in-depth investigation of packet length versus dynamics in the channel can be performed conveniently in the presented framework.

\section{Conclusions and future Work}\label{sec:conclusions}
A modular cross-layer framework for LPWANs has been presented. It allows assessing energy and reliability. Furthermore, diverse scenarios can be analyzed based on detailed energy profiles of IoT nodes. The framework has been extended with downlink messages, Adaptive Data Rate and fine-grained monitoring of various parameters (e.g., energy, collided packets). It is of particular interest to study the impact of scaling up to large numbers of nodes in a network in both quasi-static and dynamic scenarios.
Our results show that many parameters impact the energy on the link, and they do influence each other. This has been illustrated for example for the packet length and ADR parameters. The importance of a cross-layer approach is evidenced by a first specific assessment on packet length and payload size. The proposed cross-layer approaches will be validated by conducting real experiments.
%We plan to compare the presented results to measured 
We see many interesting cross-layer opportunities to further improve energy efficiency and reliability for massive Machine-Type Communication (mMTC). These include, among others, optimizing packet length taking into account channel dynamics. 

%\nocite{*}

\bibliography{bib}

\begin{thebibliography}{10}
\providecommand{\url}[1]{#1}
\csname url@samestyle\endcsname
\providecommand{\newblock}{\relax}
\providecommand{\bibinfo}[2]{#2}
\providecommand{\BIBentrySTDinterwordspacing}{\spaceskip=0pt\relax}
\providecommand{\BIBentryALTinterwordstretchfactor}{4}
\providecommand{\BIBentryALTinterwordspacing}{\spaceskip=\fontdimen2\font plus
\BIBentryALTinterwordstretchfactor\fontdimen3\font minus
  \fontdimen4\font\relax}
\providecommand{\BIBforeignlanguage}[2]{{%
\expandafter\ifx\csname l@#1\endcsname\relax
\typeout{** WARNING: IEEEtran.bst: No hyphenation pattern has been}%
\typeout{** loaded for the language `#1'. Using the pattern for}%
\typeout{** the default language instead.}%
\else
\language=\csname l@#1\endcsname
\fi
#2}}
\providecommand{\BIBdecl}{\relax}
\BIBdecl

\bibitem{semtech:lorawan_spec}
N.~SORNIN and A.~YEGIN, \emph{{LoRaWAN\texttrademark\ Specification}}, LoRa
  Alliance Technical Committee Std., 2017, v1.1.

\bibitem{Wang:2017:PNI:3070908.3071015}
\BIBentryALTinterwordspacing
Y.~P.~E. Wang, X.~Lin, A.~Adhikary, A.~Grovlen, Y.~Sui, Y.~Blankenship,
  J.~Bergman, and H.~S. Razaghi, ``{A Primer on 3GPP Narrowband Internet of
  Things},'' \emph{Comm. Mag.}, vol.~55, no.~3, pp. 117--123, Mar. 2017.
  [Online]. Available: \url{https://doi.org/10.1109/MCOM.2017.1600510CM}
\BIBentrySTDinterwordspacing

\bibitem{bor2017lora}
M.~Bor and U.~Roedig, ``{LoRa transmission parameter selection},'' in
  \emph{Proceedings of the 13th IEEE International Conference on Distributed
  Computing in Sensor Systems (DCOSS), Ottawa, ON, Canada}, 2017, pp. 5--7.

\bibitem{slabicki2018adaptive}
M.~Slabicki, G.~Premsankar, and M.~Di~Francesco, ``{Adaptive Configuration of
  LoRa Networks for Dense IoT Deployments},'' 2018.

\bibitem{8315103}
B.~Reynders, Q.~Wang, P.~Tuset-Peiro, X.~Vilajosana, and S.~Pollin,
  ``{Improving Reliability and Scalability of LoRaWANs Through Lightweight
  Scheduling},'' \emph{IEEE Internet of Things Journal}, vol.~PP, no.~99, pp.
  1--1, 2018.

\bibitem{bor2016lora}
M.~C. Bor, U.~Roedig, T.~Voigt, and J.~M. Alonso, ``{Do LoRa low-power
  wide-area networks scale?}'' in \emph{Proceedings of the 19th ACM
  International Conference on Modeling, Analysis and Simulation of Wireless and
  Mobile Systems}.\hskip 1em plus 0.5em minus 0.4em\relax ACM, 2016, pp.
  59--67.

\bibitem{pop2017does}
A.-I. Pop, U.~Raza, P.~Kulkarni, and M.~Sooriyabandara, ``{Does bidirectional
  traffic do more harm than good in LoRaWAN based LPWA networks?}'' \emph{arXiv
  preprint arXiv:1704.04174}, 2017.

\bibitem{haxhibeqiri2017lora}
J.~Haxhibeqiri, F.~Van~den Abeele, I.~Moerman, and J.~Hoebeke, ``{LoRa
  scalability: A simulation model based on interference measurements},''
  \emph{Sensors}, vol.~17, no.~6, p. 1193, 2017.

\bibitem{lorawan-efm32}
\BIBentryALTinterwordspacing
G.~Ottoy, G.~Leenders, and G.~Callebaut, ``{LoRaWAN EFM32},'' doi:
  \url{10.5281/zenodo.1209414}. [Online]. Available:
  \url{https://github.com/DRAMCO/LoRaWAN_EFM32}
\BIBentrySTDinterwordspacing

\bibitem{simulator}
\BIBentryALTinterwordspacing
G.~Callebaut, ``{LoRaWAN Network Simulator},'' doi:
  \url{10.5281/zenodo.1217124}. [Online]. Available:
  \url{https://github.com/GillesC/LoRaEnergySim/tree/v0.1.0}
\BIBentrySTDinterwordspacing

\bibitem{casals2017modeling}
L.~Casals, B.~Mir, R.~Vidal, and C.~Gomez, ``{Modeling the Energy Performance
  of LoRaWAN},'' \emph{Sensors}, vol.~17, no.~10, p. 2364, 2017.

\bibitem{semtech:modulation-basics}
\BIBentryALTinterwordspacing
\emph{{LoRa\texttrademark\ Modulation Basics}}, Semtech Std., 2015, aN1200.22.
  [Online]. Available:
  \url{https://www.semtech.com/uploads/documents/an1200.22.pdf}
\BIBentrySTDinterwordspacing

\bibitem{semtech:designers_guide}
\BIBentryALTinterwordspacing
\emph{{SX1272/3/6/7/8: LoRa Designer's Guide}}, Semtech Std., 2013, aN1200.13.
  [Online]. Available:
  \url{https://www.semtech.com/uploads/documents/LoraDesignGuide_STD.pdf}
\BIBentrySTDinterwordspacing

\bibitem{semtech:low_energy_consumption}
\BIBentryALTinterwordspacing
\emph{{SX1272/3/6/7/8: LoRa Modem Low Energy Consumption Design}}, Semtech
  Std., 2013, aN1200.17. [Online]. Available:
  \url{https://www.semtech.com/uploads/documents/LoraLowEnergyDesign_STD.pdf}
\BIBentrySTDinterwordspacing

\bibitem{petajajarvi2015coverage}
J.~Petajajarvi, K.~Mikhaylov, A.~Roivainen, T.~Hanninen, and M.~Pettissalo,
  ``{On the coverage of LPWANs: range evaluation and channel attenuation model
  for LoRa technology},'' in \emph{ITS Telecommunications (ITST), 2015 14th
  International Conference on}.\hskip 1em plus 0.5em minus 0.4em\relax IEEE,
  2015, pp. 55--59.

\bibitem{itu-r-measurement-data}
\BIBentryALTinterwordspacing
\emph{{Compilation of measurement data relating to building entry loss}},
  International Telecommunication Union Std., 2015, report ITU-R P.2346-0.
  [Online]. Available:
  \url{https://www.itu.int/dms_pub/itu-r/opb/rep/R-REP-P.2346-2015-PDF-E.pdf}
\BIBentrySTDinterwordspacing

\bibitem{damosso1999digital}
E.~Damosso, \emph{{Digital mobile radio towards future generation systems: COST
  action 231}}.\hskip 1em plus 0.5em minus 0.4em\relax European Commission,
  1999.

\bibitem{ic880a:datasheet}
\BIBentryALTinterwordspacing
\emph{{WiMOD iC880A datasheet}}, IMST Std., 2015. [Online]. Available:
  \url{https://wireless-solutions.de/downloads/Radio-Modules/iC880A/iC880A_Datasheet_V0_50.pdf}
\BIBentrySTDinterwordspacing

\bibitem{semtech:sx1301}
\emph{{SX1301 Datasheet}}, Semtech, 2017, v2.3.

\end{thebibliography}

\end{document}